\documentclass[preprint,12pt]{elsarticle}
\usepackage[utf8]{inputenc}

\usepackage{amsmath}
\usepackage{amssymb}
\usepackage{subfigure}
\usepackage{url}

\begin{document}

\begin{frontmatter}

\journal{arXiv}

\title{Data sharing games}

\author[label1,label2]{Víctor Gallego\corref{ca}}
\author[label1,label2]{Roi Naveiro}
\author[label1,label3]{David Ríos Insua}
\author[label4]{Wolfram Rozas}

\address[label1]{ICMAT-CSIC, Spain}
\address[label2]{The Statistical and Applied Mathematical Sciences Institute (SAMSI), NC, USA}
\address[label3]{School of Management, USST, China}
\address[label4]{IBM ILBD}

\cortext[ca]{Corresponding author: victor.gallego@icmat.es}

 \begin{abstract}
Data sharing issues pervade online social and economic environments. To foster social progress, it is important to develop models of the interaction 
between data producers and consumers that can promote the rise of cooperation between the involved parties.
We formalize this interaction as a game, the data sharing game, based on the Iterated Prisoner's Dilemma and 
deal with it through multi-agent reinforcement learning
techniques. 
We consider several strategies for how the citizens may behave, depending on the degree of centralization sought.
Simulations suggest mechanisms for cooperation to take place and, thus, achieve maximum social utility: 
data consumers should perform some kind of opponent modeling,
or a regulator should transfer utility between both players and incentivise them.
 \end{abstract}
 

\end{frontmatter}

\section{Introduction}

As recently discussed \citep{wolfram}, {\em data, as the intangible asset par excellence in the 21st century, is the most disputed raw material at global scale}. Ours is a data-driven society and economy,
with  data guiding most business actions and decisions. 
This is becoming even more important as
many business processes are articulated through a 
cycle of sensing-processing-acting. Indeed, Big Data is the
consequence of a digitized world where people, objects and operations are fully instrumented and interconnected, producing all sorts of data, both machine-readable (numbers and labels, known as
structured) and human-readable (text, audio or video, known as unstructured).
As data acquisition grows at sub-second speed, the capability to monetize
them arises through the ability to derive new synthetic data. 
Thus, considered as an asset, data create markets and enhance 
competition. Unfortunately, it is creating bad practices as well.
See \cite{share} for an early discussion 
as well as the 
recent European directives and legislative initiatives to 
promote public-private B2G data partnerships, e.g. \citep{europe1}.

This is the main reason for analyzing {\em data sharing 
games} with mechanisms that could foster cooperation to guarantee and promote social progress.
%
Data sharing problems have been the object of several contributions and studied
from different perspectives. For example,
 \cite{kamhoua2012game} proposes a game theoretic approach to help users determine their optimal policy in terms of sharing data in online social networks,
 based on a confrontation between a user (aimed at sharing certain information and hiding the rest) and an attacker (aimed at exposing the user's PI
 or concealing the information the user is willing to share). This
  is modelled through a zero-sum Markov game; a  Markov equilibrium is computed and the corresponding Markov strategies are used to give advice to users.
 \cite{figueiredo2017data} reviews the impact of data sharing in science and society and presents guidelines to improve the efficiency of 
 data sharing processes,
 quoting \cite{pronk}, who provide a game theoretical analysis suggesting 
 that sharing data with the community can be the most profitable and stable strategy. Similarly, \cite{dehez2013data} consider a setting in which 
 a group of firms must decide whether to cooperate in 
 a project that requires the combination of data held by several of them; the authors  address the question of how to compensate the firms for the data they
 contribute with,
 framing the problem as a transferable utility game and characterizing its Shapley value as a compensation mechanism. 
  
Our approach models interactions between data owners and consumers 
inspired by the iterated prisoner's dilemma (IPD) \cite{axelrod84}. This is an elegant incarnation of the problem of how to achieve agents` cooperation in competitive settings. Other authors have used similar models in other socio-technical problems as in politics 
\citep{brams}, and security \citep{kunreuther}, among others. Our approach to model agent's behavior is different and relies on multi-agent reinforcement learning (MARL) arguments \citep{gallego2019opponent}. Reinforcement learning (RL)has been successfuly applied to games that are repeated over time, thus making it possible for agents to optimize their strategies \citep{LAHKAR201310}. The work of \cite{SHAFRAN2012354} also discusses the use of RL in iterated games such as Prisoner's Dilemma, although they do not focus on the issue of incentivizing cooperation between the players. Through RL we are able 
to identify relevant mechanisms to promote cooperation.

The structure of the paper is as follows. First, a qualitative description of the problem, the intervening agents and their strategies is provided. 
We next model it quantitatively and develop scenarios 
that could promote cooperation through MARL. 
We study those in simulated environments confirming that cooperation is both possible and the best social strategy, ending with a brief discussion.

\section{Data sharing: categories, agents and strategies}
Before modeling interactions between data consumers and producers, it is convenient to understand the data categories available. 
Even though admittedly with a blurry frontier, 
from a legal standpoint, there are two main ones:

\begin{itemize}
\item {\em Data that should not be bought/sold}. This refers to  
 personal information (PI), as e.g.\ the data preserved in the European
 Union through 
 the General 
 Data Protection Regulation (GDPR) \citep{gdpr} and other citizen defense frameworks
 aimed at guaranteeing civic liberties. PI includes 
 data categories such as 
{\em internal information} (like knowledge and beliefs, 
 and health 
 data); 
{\em financial information} (like accounts or 
 credit data);
{\em social information} (like 
criminal records 
or communication data); or,
{\em tracking information} (like computer device; 
or location data).
\item {\em Data that might be purchased}. Citizen’s data is a property,
there being a need to guarantee a fair and transparent compensation.
Accountability mitigates market frictions. For
traceability and transparency reasons,
blockchain-based platforms are being implemented at the moment.
\end{itemize}
  A characterization of what type of data 
  belongs to each category will depend on the context and is, most of the times, subjective.

In any case, in the last decades, modern data analytics techniques and strategies are enabling the generation of new types of data:
\begin{itemize}
\item {\em Data that might be estimated/derived.}  Currently available analytics technologies have the ability of estimating efficiently citizen behavior and other characteristics by deeply analyzing Big Data. For instance, platforms such as IBM Personality Insights \citep{ibm} can estimate personality traits of a
given individual using his/her tweets, thus facilitating marketing activities.
As a result, the originating data becomes a new asset for a company
willing to undertake its analysis.
\end{itemize}

Having mapped the available data, there is a need to understand the 
knowledge actually available and how is it uncovered.	
Within the above scenario, 
we consider two players in  a data 
sharing game: the data providers 
(Citizen, she) and the Dominant Data Owner (DDO, he).
A DDO could be a private
company, e.g. GAFA (Google, Apple, Facebook, Amazon) or Telefonica,
or a public institution (Government). 
Inspired by the classic Johari window \citep{johari},
we inter-relate now what a Citizen knows or does not
with what a DDO knows or does not
to obtain these scenarios:
\begin{enumerate}
\item Citizen knows what DDO does.
The citizen has created 
a data asset which she sells to a DDO. 
Sellable data create a market which could 
evolve in a sustainable manner if accountability and transparency are somehow guaranteed.
\item 	Citizen knows what DDO does not.
This is the PI realm.
Citizens would want legal frameworks like the 
GDPR or data standards preserving  
citizen rights, mainly 
ARCO-PL (access, rectification, cancellation, objection, portability
and limitation) so that PI is respected.
\item Citizen does not know what DDO does. 
The DDO has unveiled
citizen’s PI through deep analysis of
Big Data.\footnote{As in the famous Target pregnant
teenager case \cite{target}} 
This analysis may be acceptable if data are dealt just as a target.
Data protection frameworks should guarantee civil rights and liberties in such activities.
\end{enumerate}
Note that we could also think of a fourth scenario in which 
neither the citizen knows, nor the DDO does, although this is clearly unreachable. 

Once explained how knowledge is shared, we analyze how 
knowledge creation can be fostered to stimulate social progress, studying 
cooperation  scenarios between Citizen and DDO. 
We simplify by considering two strategies
for both players, respectively designated {\em Cooperate}  (C)
and {\em Defect} (D), leading to the four scenarios 
in Table
\ref{kaka}.

{\small
\begin{table}[htbp]
	\centering
	\scalebox{0.8}{
	\begin{tabular}{c|c|c}
			        &  DDO cooperates & DDO defects  \\
		\hline  
Citizen cooperates &  Citizen sells data, &Citizen taken for a ride\\
	     	&  demands data protection &  selling data, while DDO                 \\
   	        &  DDO 	purchases and  &    does not pay Citizen         \\
   	        &  respect Citizen data.    &   data with services            \\ \hline
Citizen defects  &    DDO  taken for a ride & Citizen sells wrong/noisy  \\ 
		  &    purchasing. Citizen     & data does not pay for DDO  \\
		  &  selling   wrong/noisy  &  services, who does not pay data  \\
		  &  data becomes free rider.         & with  services.  \\ 		
			\end{tabular}%
			}
	\caption{Scenarios in the data sharing game.}
	\label{kaka}%
\end{table}
}

Reflecting about them, the only one 
that ultimately fosters knowledge creation and, therefore, stimulates social progress, is mutual cooperation. It is the best scenario and produces mutual value. Cooperation begs for a FATE (fair, accountable, transparent, ethical) technology like blockchain. In such scenario, data (Big Data), algorithms and processing technology would boost knowledge. Mutual cooperation is underpinned by decency and indulgence values 
such as being
{\em nice} (cooperate when the other party does); 
{\em provokable} (punish non cooperation);
{\em forgiving} (after punishing, immediately cooperate, reset credit);
and {\em clear} (the other party easily understands and realises that 
the best next move is to cooperate).

Mutual defection is the worst scenario 
in societal terms: it produces a data market failure, stagnating social progress. As there is no respect from both sides, no valuable data trade will happen, 
and even a noisy data vs.\ unveiled data war will take place. Loss of freedom may arise as a result. 

The scenario (Citizen cooperates, DDO defects) is the worst
for the citizen, leading to data power abuses, as with the
 UK ``ghost" plan. 
It would generate asymmetric information,
adverse selection, and moral hazard problems, in turn producing 
data market failures.
The DDO behaves incorrectly, there being a need to punish unethical
and illegal behaviour. As an example, the GDPR sets the right to receive explanations for
algorithmic decisions. There is also a need for mitigating
systematic cognitive biases in algorithms. Citizens may respond by 
sending noisy data,
rejecting data services, imposing standards over data services or setting prices 
according to success. 

Finally, the scenario (Citizen defects, DDO cooperates) is the worst for the DDO. It leads to data market failures and shrinks knowledge. This  
stems from  a behavior of not paying for
public/private services that can be obtained anyway. 
In the long run, this erodes public and private services quality and creativity. This misbehavior should be punished to restore cooperation and a fair price should be demanded for services.

\section{A model for the data sharing game}\label{sec:models}

We model interactions between citizens and DDOs over 
time from the perspective of the IPD. 
Table \ref{tab:payoffIPD} shows its reward bimatrix. 
The row player will be the Citizen, for whom {\em cooperate} means that she
wishes to sell and protect her data, whereas {\em defect} means she either sells wrong data or decides not to contribute. The DDO will be the column player
for whom {\em cooperate} means that he purchases and protects data,
whereas {\em defect} means that he is not going to pay for the collected data or will not protect it.
Payoffs satisfy the usual conditions in the IPD, that is $T>R>P>S$ and $2R > T+S$.
When numerics are necessary,
we adopt the choice $T= 6 $, $R= 5 $, $ P = 1 $,  and $S= 0 $.

\begin{table}[]
\begin{center}
\begin{tabular}{cl|lll}
\multicolumn{1}{l}{}                                   &     & \multicolumn{3}{l}{\textbf{DDO}} \\ \cline{3-5} 
\multicolumn{1}{l}{}                                   &     & $C$         &       & $D$        \\ \hline
\multicolumn{1}{c|}{\textbf{Citizen}} & $C$ & $R,R$       &       & $S,T$      \\
\multicolumn{1}{c|}{}                                  &     &             &       &            \\
\multicolumn{1}{c|}{}                                  & $D$ & $T,S$       &       & $P,P$     
\end{tabular} 
\end{center}
\caption{Payoffs in the data sharing game}
\label{tab:payoffIPD}
\vspace{-2ex}
\end{table}

It is well-known that in the one-shot version of the IPD game, the unique Nash equilibrium is $(D,D)$, leading to the social dilemma described above: the selfish rational point of view of both players leads to an inferior societal position. Similarly, if the game is played $N$ times,
and this is known by the players, these have no incentive to cooperate, as we may reason by backwards induction \citep{axelrod81}. 
However, in realistic scenarios, players are not sure about 
whether they will meet 
again in future and, consequently, they cannot be sure when the last interaction will be taking place \citep{axelrod84}. Thus, it seems reasonable to assume that players will interact an indefinite number of times or that there is a positive probability of meeting again. This possibility that players might interact again is precisely what makes cooperation emerge.





The framework that we adopt to deal with the problem is MARL 
\citep{marl_over}. Each agent $a \in \lbrace C, DDO \rbrace $ maintains its policy $\pi_a(d_a|o_a, \theta_a)$ used to select a decision $d_a$ under some observed state of the game $o_a$ (for example, the previous pair of decisions) and parameterised by
certain parameters $\theta_a$. Each agent learns
how to make decisions by optimizing his policy under the expected sum of discounted utilities
$$
\max_{\theta_a} \mathbb{E}_{\pi_a} \left[ \sum_{t=0}^\infty \gamma^t r_{a, t} \right],
$$
where $\gamma \in (0,1)$ is a discount factor and $r_{a,t}$ is the reward that
agent $a$ attains at time $t$.
The previous optimization can be performed through 
Q-learning or policy gradient methods \citep{sutton2012reinforcement}. 
The main limitation with this approach in the multi-agent setting is that if the
agents are unaware of each other, they are shown to fail to cooperate 
\citep{gallego2019opponent}, leading to defection every time, which is undesirable in the data sharing  game.

As an alternative, in order to foster collaboration, we propose three approaches, depending on the degree of decentralization and incentivisation sought for.
\begin{itemize}
        \item In a (totally) decentralized case, C and DDO are alone and we resort to opponent modelling strategies,
        as showcased in Section \ref{sec:decentralized}. However, 
        this approach may fail under severe misspecification in the opponent's model. Ideally, we would
        like to encourage collaboration without making strong assumptions about learning algorithms used by each player. 
      
        \item Alternatively, a third-party could become a regulator of the data market: C and DDO use it and  
        the regulator introduces taxes, as showcased 
        in Section \ref{sec:regulator}. 
        The benefit of this approach is that the regulator only needs to observe the actions adopted by the agents, not needing to make any assumption about their models or motivations and optimizing their behaviors based on whatever social metric he considers.
        
        \item Finally, in Section \ref{sec:incentives} we augment the capabilities of the previous regulator to enable it to  incentivize the agents, leading to further increases in the social metric considered.
    \end{itemize}

\noindent To fix ideas, we focus on a social utility (SU) metric
defined as the agents' average utility  
\begin{equation}\label{eq:su}
SU_t = \frac{r_{C,t} + r_{DDO,t}}{2}.
\end{equation}
This requires adopting a notion of transferable utility,
serving as a common medium of exchange that can be transferred between agents, see e.g. \citep{aumann1960}.


\section{Three solutions via Reinforcement Learning}

\subsection{The decentralized case}\label{sec:decentralized}
Our first approach models the interaction between both agents as an IPD, and simulates such interactions to assess the impact of different 
DDO strategies over social utility. We first fix the strategy of the DDO, assume that the citizen models the DDO behaviour and simulate interactions between
both agents 
finally assessing social utility.\footnote{Code for all the simulations performed can be found at \url{https://github.com/vicgalle/data-sharing}}

We  model the Citizen as a Fictitious Play Q-learner (FPQ) in the spirit of \cite{gallego2019opponent}.
She chooses her action $d_a \in \lbrace C, D \rbrace$
maximizing her expected utility $\psi(d_a)$
defined through 
\[ \psi(d_a) = \mathbb{E}_{p_{FP}(d_b)} [Q(d_a,d_b)] = \sum_{d_b \in \lbrace C, D \rbrace } Q(d_a, d_b) p_{FP}(d_b), \]
 where $p_{FP} (d_b)$ reflects the Citizen's beliefs about her opponent's
 actions $d_b \in \lbrace C, D \rbrace$
 and 
 $Q(d_a,d_b)$ is the augmented Q-function from the threatened Markov decision processes as defined in \cite{gallego2019opponent}, 
 an estimate of the expected utility obtained by the Citizen if both
 players were to commit to actions $d_a, d_b$.
 
 We estimate the probabilities $p_{FP} (d_b)$
  using the empirical frequencies of the opponent's past
  plays as in Fictitious Play 
 \cite{brown1951iterative}. To further favor learning, the Citizen could 
place a Beta prior over $p_C \sim \mathcal{B}(\alpha, \beta)$,
the probability of the DDO cooperating, 
with probability $p_D = 1-p_C$ of defecting.
 Then, if the opponent chooses,
for instance, {\em cooperate}, the citizen updates her
beliefs leading to the posterior $p_C \sim \mathcal{B}(\alpha + 1, \beta)$, 
and so on. 

We may also augment the citizen model to have memory
of the previous opponent's action. This can be
straightforwardly done replacing $Q(d_a,d_b)$ with $Q(s,d_a,d_b)$ and $p_{FP}(d_b)$ with $p_{FP}(d_b|s)$
where $s \in \lbrace C, D \rbrace \times \lbrace C, D \rbrace$ is
the previous pair of actions both players took. Thus, we  
need to keep track of four Beta distributions, one for each
value of $s$. This FPQ agent with
memory will be called FPM. 
Clearly, this approach could be expanded to account for 
longer memories over the action sequences. However,  \cite{press2012iterated} shows that agents with a 
good memory-1 strategy can effectively force the iterated 
game to be played as memory-1, even if the opponent has a
longer memory.

\subsubsection{Experiments}

We simulate the previous IPD under different strategies 
for the DDO and measure the impact over social utility. For each scheme, we display the social utility  attained over time by the agents. For all experiments, we model the citizen as an FPM agent (with memory-1). The discount factor was set to 0.96. 

\begin{figure*}[h!]
\centering
\subfigure[Agents' utilities.]{%
  \label{fig:nash_ut}%
  \includegraphics[height=1.8in]{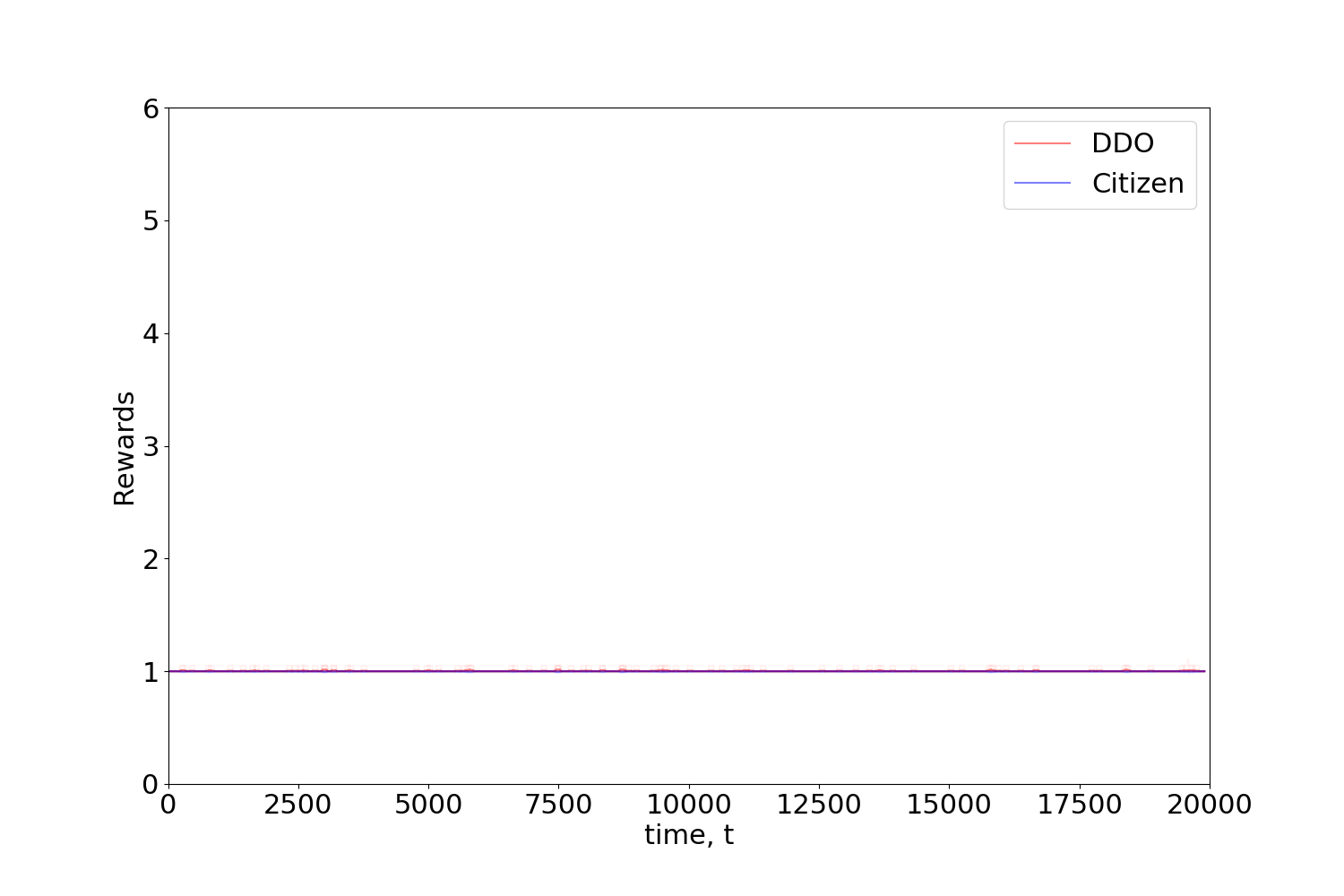}}%
  \subfigure[Social utility.]{%
  \label{fig:nash_sut}%
  \includegraphics[height=1.8in]{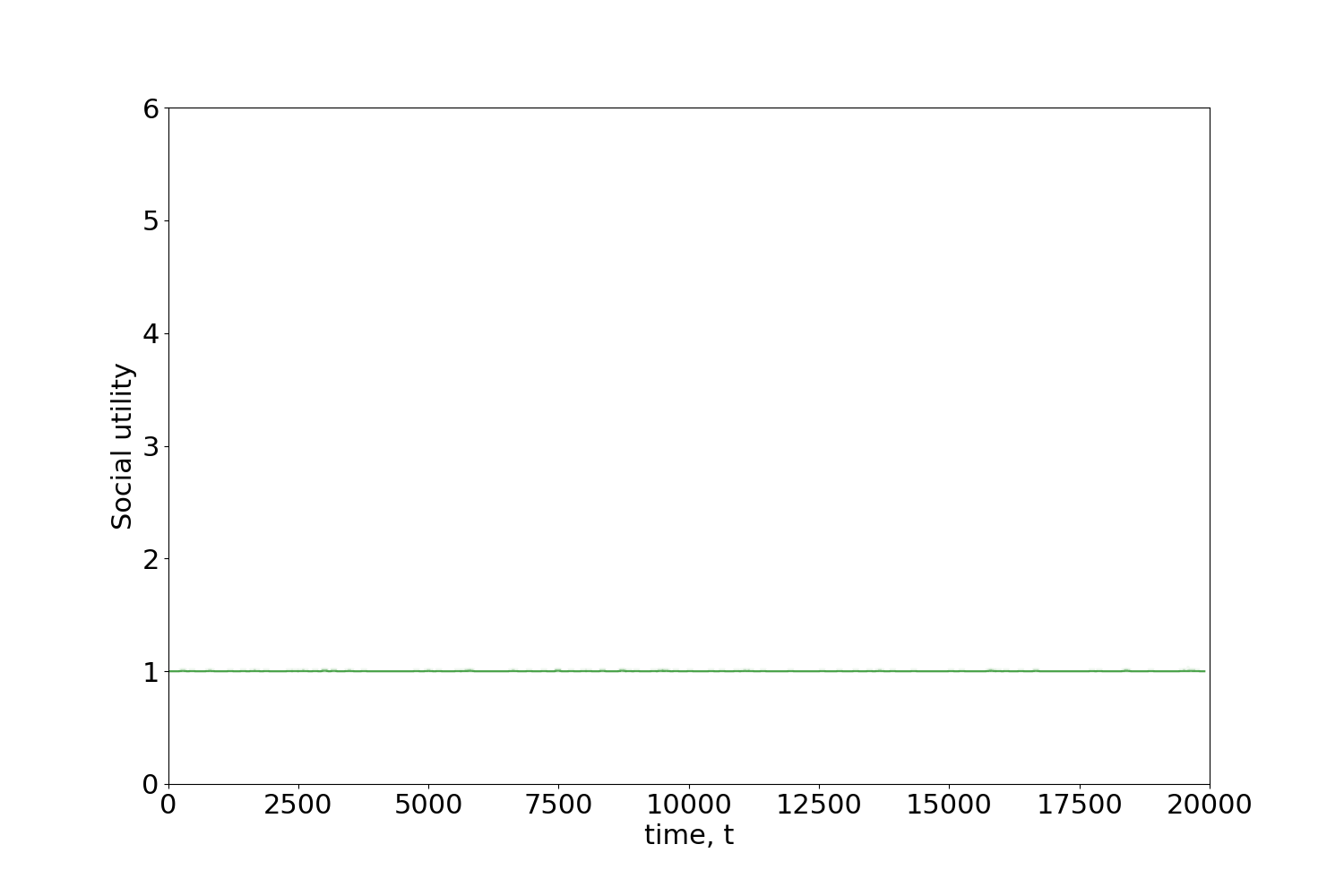}}%
  \caption{Agents' utilities and social utilities in case of DDO always defecting.}
\end{figure*}

\paragraph{Selfish DDO} 
When we assume a selfish DDO, playing always defect, our simulation confirms that this strategy will force 
the citizen to play defect and sell wrong data, not having incentives to abandon
such strategy. 
Even when 
citizens have strong prior beliefs that the DDO will cooperate, after a
few iterations they will learn that the DDO is defecting always and thus choose also to defect, as shown in Figure \ref{fig:nash_ut}.

\noindent Figure \ref{fig:nash_sut} shows that under the defecting strategy,
the social utility achieves its minimum value. 

\begin{figure}[h!]
\centering
\includegraphics[width=0.6\linewidth]{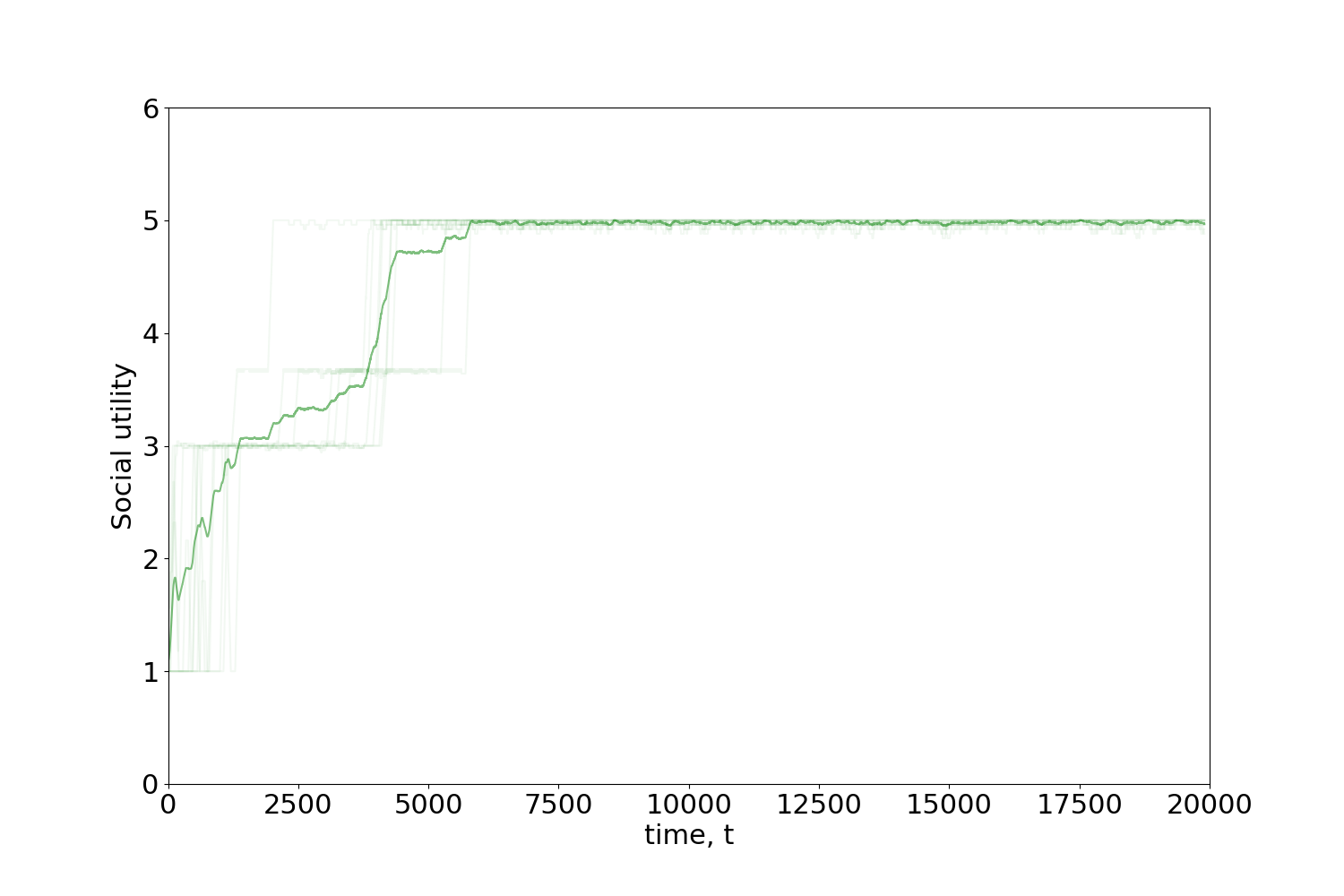}%
\caption{Social utility of a FPM citizen against a TfT DDO.}\label{fig:FPMvsTfT}
\end{figure}

\paragraph{A Tit for Tat DDO}
We next model the DDO as a player using the Tit for Tat (TfT) strategy
(it will first cooperate and, then, subsequently replicate the opponent's previous action: if the opponent was previously cooperative, the agent is cooperative; if not, it defects).
This policy has been widely used in the IPD, because of its simplicity and effectiveness \cite{axelrod84}. A recent experimental study 
\cite{dal2019strategy}
 tested real-life people's behaviour in IPD scenarios,
 showing that TfT was one of the most widely strategies.
 Figure \ref{fig:FPMvsTfT} shows that under TfT, the 
 social utility achieves its maximum value: mutual cooperation is achieved, thus leading to the optimal social utility.

It is important to mention though that 
if the citizen had no memory about previous actions, the policy of the DDO could not be learnt and mutual cooperation would not be achieved.

\begin{figure}[h!]
\centering
\includegraphics[width=0.6\linewidth]{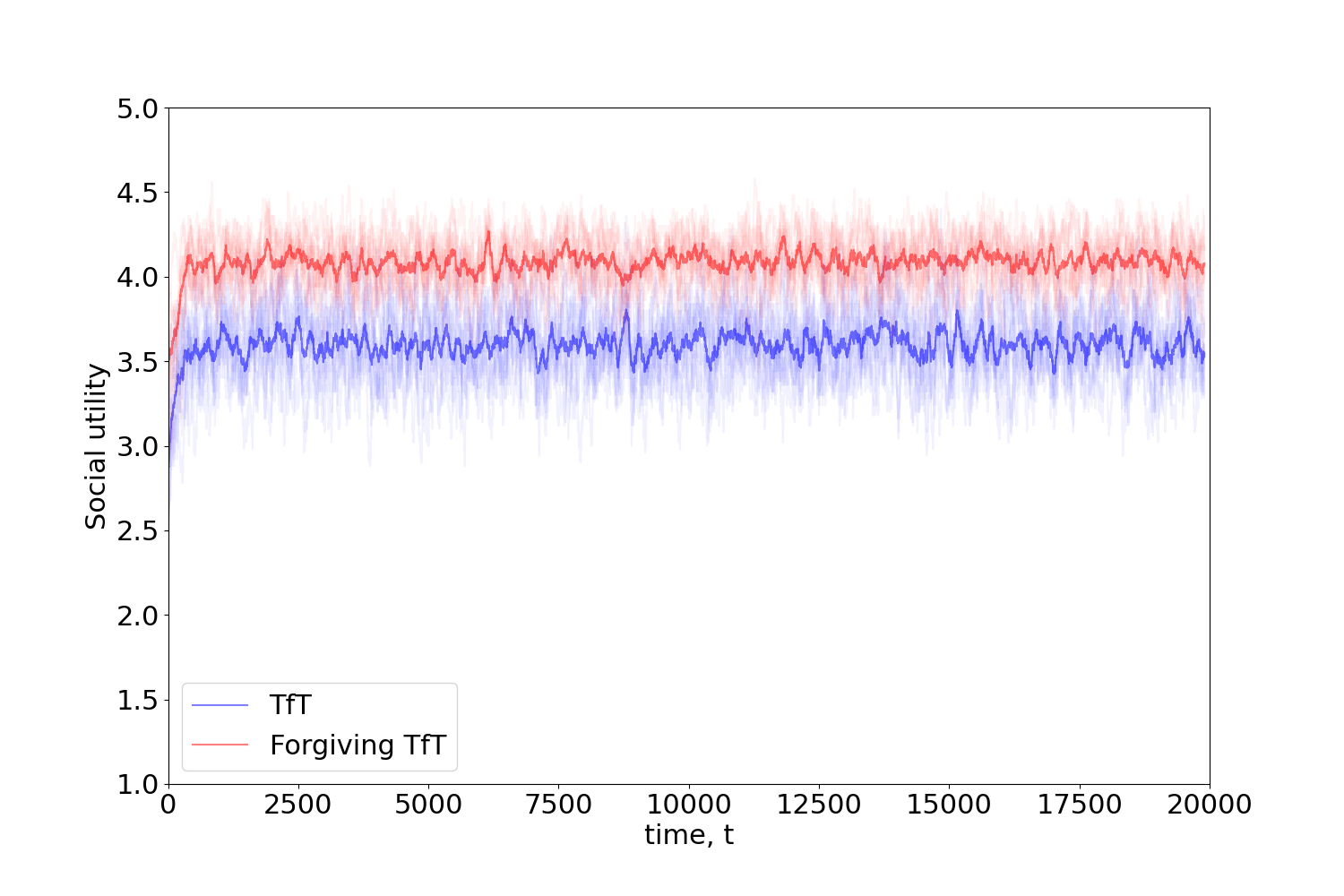}%
\caption{Social utility when citizens act randomly $70$ \% of the time against a TfT and a forgiving TfT DDO.}\label{fig:25}
\end{figure}

\paragraph{Random behaviour among citizens}
 Previously, we all citizens were assumed to
 act according to the FPM model.
 However, assuming that the whole population will behave following such complex strategies is unrealistic. A more reasonable
 assumption considers having a  subpopulation of citizens that acts randomly. To simulate this, we modify the FP/FPM model to draw a random action with probability $0 < \epsilon < 1$ at each turn. As Figure \ref{fig:25} shows, where we set $\epsilon = 0.7$, this entails a huge decrease in social utility. 


\paragraph{A forgiving DDO}
A possible solution for this decrease in social utility consists of forcing the DDO to eventually forgive the Citizen and play cooperate regardless of
her previous actions. We model this as follows: with probability $p$ the DDO will cooperate, whereas with probability $1-p$ he will play TFT. 
%
%

To assess what proportion of time should the DDO forgive, we evaluated a grid of values from 0 to 100, and chose the one that produced the highest increase in social utility. The optimal value was forgiving $70 \%$ of times.
As Figure \ref{fig:25} shows, this produces an increase of approximately half a unit in the average social utility with respect to the case of never forgiving.

Note, though, that there exists a limit value for the forgiving rate such that, if surpassed, the social utility will decrease to around 3. The reason for this is that, in this regime, when not acting randomly, the Citizen will learn that the DDO cooperates most of the time, and thus her  optimal strategy will be to defect. Thus, in most iterations the actions chosen will be $(C, D)$, leading to a social utility of around 3.



\subsection{Taxation through a regulator}\label{sec:regulator}

We discuss now an alternative solution to promote cooperation introducing a third player, a Regulator (R, it). Its objective is to nudge the behaviour of the other players through utility transfer, based on taxes.   \ref{one-shot}
discusses a one-shot version identifying its equilibria.
As in Section 4.1, our focus 
is on the iterated version of this game.

At each turn, the regulator will choose a tax policy for the agents 
$$
(tax_{C, t}, tax_{DDO,t}) \sim \pi_R(\cdot | o_R, \theta_R), 
$$
where $o_R$ is the observed state of the game
and $\theta_R$ are relevant parameters for the regulator. 
Then, the other two agents will receive their corresponding adjusted utility $\tilde{r}_{a,t}$ through 
$$
\tilde{r}_{a,t} = r_{a, t} - tax_{a,t} + \frac{1}{2} \sum_a tax_{a, t},
$$
where the first term is the original utility (Table \ref{tab:payoffIPD});
the second one is the tax that the regulator collects 
from that
agent; and, finally, the third one is is the (evenly) redistributed collected 
reward.  Note that
$$
SU_t = \frac{r_{C, t} + r_{DDO, t}}{2} = \frac{\tilde{r}_{C, t} + \tilde{r}_{DDO, t}}{2}. 
$$
Thus, under this new reward regime, utility is not created nor destroyed, only transferred between players.

Let us focus now on the issue of how does the Regulator learn its
tax policy. For this, we make it another RL agent that maximizes the social welfare function,
thus optimizing its policy by solving 
$$
\max_{\theta_R} \mathbb{E}_{\pi_R} \left[ \sum_{t=0}^\infty \gamma^t SU_t \right].
$$
Then, two nested RL problems are considered: first, 
the regulator selects a tax regime and, next, the other two players optimally adjust their behaviour to this regime. After a few steps, 
the regulator updates its
policy to further encourage cooperation (higher $SU_t$), and so on. At the end of this process, we would expect that both players' behaviours would have been nudged towards cooperation.
 
 We thus frame learning as a bi-level RL problem with two nested loops, 
 using policy gradient methods:
\begin{enumerate}
    \item \textbf{(Outer loop)} The regulator 
    has parameters $\theta_R$, imposing a certain tax policy.
    \begin{enumerate}
        \item \textbf{(Inner loop)} The agents learn under this tax policy for $T$ iterations:
        \item They update their parameters: $\theta_{a, t+1} = \theta_{a, t} + \eta \nabla \mathbb{E}_{\pi_a} \left[ \sum_{t=0}^\infty \gamma^t r_{a, t} \right] $.
    \end{enumerate}
    \item The regulator updates its parameters: $\theta_{R, t+1} = \theta_{R, t} + \eta \nabla \mathbb{E}_{\pi_R} \left[ \sum_{t=0}^\infty \gamma^t SU_t \right]  $.
\end{enumerate}


Let us highlight a few benefits of this approach.
First, the Regulator makes no assumptions about the policy models of the other players (thus it does not matter whether they are just single-RL agents or are opponent-modelling). Moreover,
this framework is also agnostic to the \emph{social welfare function} to
be optimized; for simplicity, we just use the expression (\ref{eq:su}).
     It is also scalable to more than two players: 
     the regulator only needs to collect taxes for each player, and then redistributes wealth.
   In case that we were considering $k > 2$ agents, we would have to split the sum of taxes by $1/k$. 

\subsubsection{Experiments}

This experiment 
illustrates the performance of the general 
framework, showing how the inclusion of a Regulator encourages the emergence of cooperative behavior.

Consider the interactions between a Citizen and a DDO. 
The parameter for 
each player is a vector $\theta_a \in \mathbb{R}^2$, with $a \in \lbrace C, DDO\rbrace $, representing the logits of choosing the actions,
i.e. the unnormalized probabilities of choosing each decision.
We consider two types of regulators.

The first one has a discrete action space defined through
$$
        tax_{a,t} =  \begin{cases} 0.00 & \text{if } a_R = 0\\
        0.15\cdot r_{a,t} & \text{if } a_R = 1\\
        0.30\cdot r_{a,t} & \text{if } a_R = 2\\
        0.50 \cdot r_{a,t} & \text{if } a_R = 3. \end{cases}
$$
For example, when $a_R=2$ the tax rate reaches 30\%.
In this case, $\theta_R \in \mathbb{R}^4$ represent the 
logits of a categorical 
random variable taking the previous values (0,1,2,3).

The second regulator adopts a Gaussian policy defined 
through $\pi_R(d_R | o_R, \theta_R) \sim \mathcal{N}(d_R | \theta_R, 0.05^2)$, with tax 
$$
tax_{a,t} = 0.5 \cdot sigmoid(d_R) \cdot r_{a,t},
$$
to allow 
for a continuous range in $\left[ 0, 0.5 \right]$.

\begin{figure}[!h]
\centering
\includegraphics[scale=0.7]{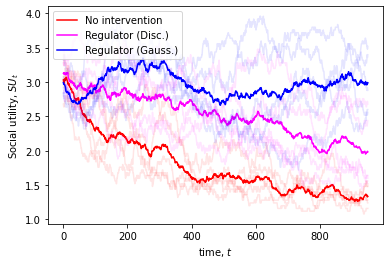}
\caption{Social utility under three different regulation scenarios.}\label{fig:su}
\end{figure}

Experiments run for $T=1000$ iterations.
After each iteration, both agents perform one update of
their policy parameter gradient.
The regulator updates its parameters using policy gradients every 50 iterations.
The decision of updating the regulator less frequently than the other agents is motivated to allow them to learn and adapt to the new tax regime and stabilise
the overall learning of the system.
Figure \ref{fig:su} displays results. 
 For each of the three variants (no intervention, discrete, continuous) we plot 5 different runs and their corresponding means in darker color.
 
 Clearly, under no intervention, both agents fail to learn to cooperate
converging to the static 
Nash equilibrium $(D,D)$. 
We also appreciate that the discrete policy is neither 
effective, also converging to $(D,D)$, albeit at a much slower pace.
On the other hand, the Gaussian regulator is 
 more efficient as it allows to avoid convergence to $(D,D)$
although it does not preclude convergence to $(C,C)$.
This regulator is more effective than its discrete counterpart,
because it can better exploit 
the policy gradient information. Because of this, in the next subsection we will  focus on this Gaussian regulator.

In summary, the addition of a Regulator can make a positive impact in the social utility attained in the market, preventing collapse into $(D, D)$. However, introducing taxes to the players is not sufficient, since in Figure \ref{fig:su} the social utility converged towards a value of 3, far away from the optimal value of 5.

\subsection{Introducing incentives}\label{sec:incentives}

In order to further stimulate cooperative behavior, we introduce incentives to the players via the Regulator: if both players cooperate at a given turn, they will receive an extra amount $I$ of utility,
 a scalar that adds to their perceived rewards. \ref{sec:oneshot_inc} shows that incentives complement well with the tax framework, so that mutual cooperation is possible in the one-shot version of this game. Note that, when $I>T-R$, instead of the Prisoner's Dilemma, we have an instance of the Stag Hunt game \cite{skyrms2004stag}, in which both $(C, C)$ and $(D, D)$ are 
 pure Nash equilibria.\footnote{Achieving mutual cooperation is much simpler in this case.}

\begin{figure}[!h]
\centering
\includegraphics[scale=0.7]{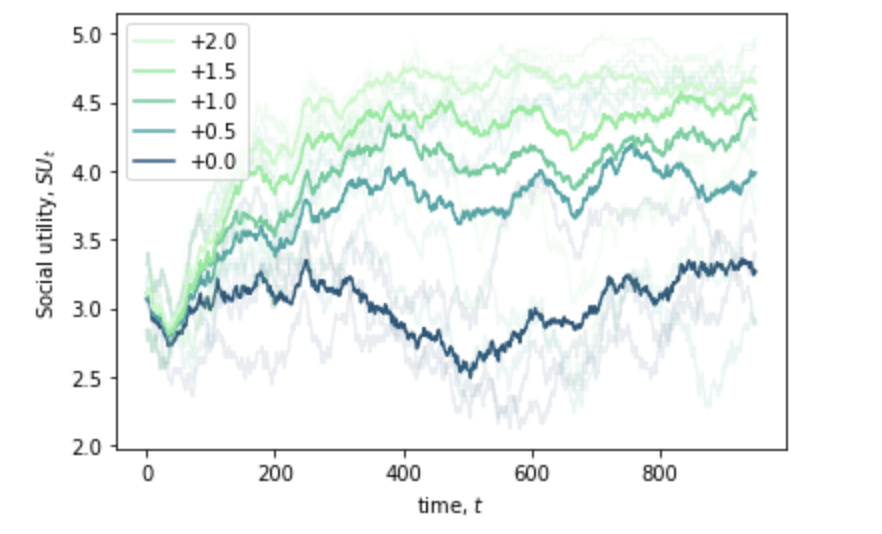}
\caption{Social utility under different incentives with tax collection.}\label{fig:inc1}
\end{figure}

From now on, we focus the discussion in the iterated version.
In this batch of experiments, players interact over $T=1000$ iterations, and the Regulator only provides incentives during the first 500 iterations. After that, he will only collect taxes from the players and redistribute them as in Section \ref{sec:regulator}. Figure \ref{fig:inc1} shows
results from 
several runs under different incentive values. A few comments are in order.

\begin{figure}[!h]
\centering
\includegraphics[scale=0.7]{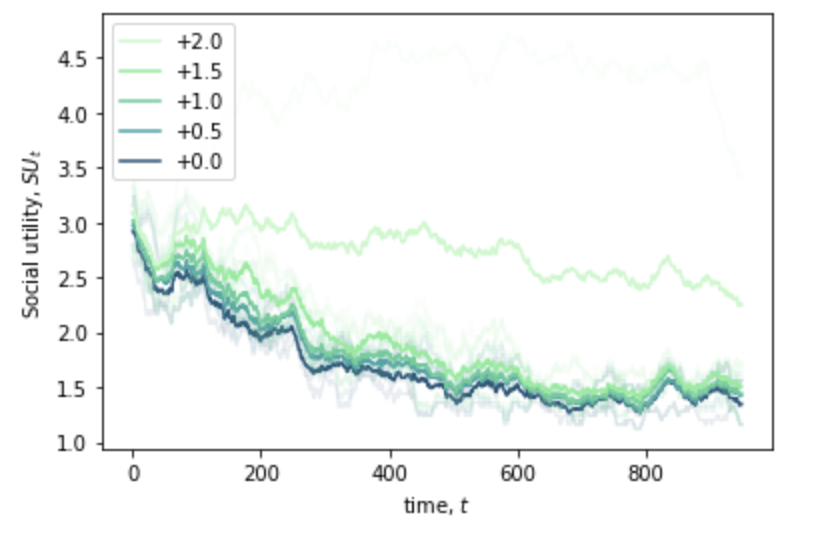}
\caption{Social utility under different incentives with no tax collection.}\label{fig:inc2}
\end{figure}

Firstly, note that as the incentive increases, also does the social utility.
For an incentive of 1, the maximum reward of $(C, C)$ and $(D, C)$ is the same (6) for the Citizen, and cooperation emerges naturally. Also note that since the policies for each player are stochastic, it is virtually impossible to maintain an exact convergence towards the optimal value of 5, since a small amount of time the agents are deviating from $(C,C)$ due to the stochasticity in their actions. Second,
observe that even when the Regulator stops incentivizing players in the middle of the simulations, both players keep cooperating along time.

We hypothesize that the underlying tax system from Section \ref{sec:regulator} is necessary for players to learn to cooperate and maintain that behaviour even after the Regulator stops incentivizing them. To test this hypothesis, we repeat the experiments removing tax collection, \emph{ceteris paribus}. Results are shown in Figure \ref{fig:inc2}. Observe now that even under the presence of high incentives, both agents fail to cooperate, with social utility decaying over time. Thus, we have shown that the tax collection framework from \ref{sec:regulator} has a synergic effect with the incentives introduced in this Section.

\section{Discussion}
A defining trend in modern society is the abundance 
of data which opens up new opportunities, challenges
and threats. In the upcoming years, social progress will be essentially conditioned by the capacity of society to gather, analyze and understand data, as this will 
facilitate better and more informed decisions. 
Thus, to guarantee social progress,  efficient mechanisms
for data sharing are key. Obviously, such mechanisms
should not only facilitate the data sharing process, 
but must also guarantee the protection of the citizen's personal
information. As a consequence, the problem of data sharing not only
has importance from a socioeconomic perspective, but also from the
legislative point of view. This is well described in numerous recent legislative pieces from the 
EU, e.g.\ \cite{europe1}, as well as in the concept of flourishing 
in a data-enabled society \citep{allea}. 


We have studied the problem of data sharing
from a game theoretic perspective with two agents.
Within our setting,  mutual cooperation emerges as the strategy 
leading to the best social outcome, and it must be promoted somehow. We
have proposed modelling the confrontation between dominant data owners and citizens using two versions of the iterated prisoner dilemma via  
multi agent reinforcement learning: the decentralized case, in which both agents interact freely, and the centralized case, in which the interaction is regulated by an external agent/institution. In the first case, we have shown that there are strategies with which mutual cooperation is possible, and that a forgiving policy by the DDO can be beneficial in terms of social utility. In the centralized case, regulating the interaction between citizens and DDOs via an external agent could foster mutual cooperation through taxes and incentives.

 Besides fostering cooperation,  the data sharing game may be seen as
  an instance of a two sided market \citep{rochet2006two}. Therefore,
  the creation of intermediary platforms that facilitate the connection between dominant data owners and citizens to enable data sharing
  would be key to guarantee social progress. 

\paragraph*{Acknowledgements}
This work was partially supported by the NSF under Grant DMS-1638521 to the Statistical and Applied Mathematical Sciences Institute
and a BBVA Foundation project.
RN also acknowledges support of the Spanish Ministry for his grant FPU15-03636.
VG also acknowledges support of the Spanish Ministry for his grant FPU16-05034.
DRI is grateful to the MTM2017-86875-C3-1-R AEI/ FEDER EU project,
and the AXA-ICMAT Chair in Adversarial Risk Analysis.

\section*{References}
\bibliographystyle{plainurl}
\bibliography{references}

\pagebreak

\appendix
\section{One-shot game for the centralized case}\label{one-shot}

    


We model the one-shot version of the centralized case game as a
three-agent sequential game.
The regulator acts first choosing a tax policy; 
after observing it, the agents take their actions. 
Introducing a regulator can foster cooperation in the one shot game.

For simplicity, consider the following policy: the regulator will retain a percentage $x$ of the reward if the agent decides to defect, and 0 if it decides to cooperate. Then, the regulator will share evenly the amount collected between 
both agents. With this, given the regulator's action $x$, the payoff matrix 
is as in Table \ref{tab:payoffIPD2}, recalling that  $T > R > P >S$.
    
    \begin{table}[h!]
    \begin{center}
    \begin{tabular}{cl|lll}
    \multicolumn{1}{l}{}                                   &     & \multicolumn{3}{l}{\textbf{DDO}} \\ \cline{3-5} 
    \multicolumn{1}{l}{}                                   &     & $C$         &       & $D$        \\ \hline
    \multicolumn{1}{c|}{\textbf{Citizen}} & $C$ & $R,R$       &       & $S',T'$      \\
    \multicolumn{1}{c|}{}                                  &     &             &       &            \\
    \multicolumn{1}{c|}{}                                  & $D$ & $T',S'$       &       & $P,P$     
    \end{tabular} 
    \end{center}
    \caption{Utilities for the data sharing game}
    \label{tab:payoffIPD2}
    \vspace{-2ex}
    \end{table}
    
    Assume that if one agent defects and the other cooperates, the first one will receive a higher payoff, that is $T' > S'$, which means that $x < 1 - \frac{S}{T}$. Depending on $x$, three scenarios arise:
    \begin{enumerate}
        \item $T' > R > P > S' \iff x < 2 \left[ \frac{P-S}{T}\right]$. This is
        equivalent to the prisoner's dilemma. $(D,D)$ strictly dominates, thus being the unique Nash Equilibrium. 
        
        \item $R > T' > S' > P \iff x > 2 \left[ 1- \frac{R}{T}\right]$. In this case, $(C,C)$ strictly dominates, becoming the unique Nash Equilibrium. 
        
        \item $T' > R > S' > P \iff x \in \left( 2 \left[ \frac{P-S}{T}\right], 2 \left[ 1- \frac{R}{T}\right] \right) $. This is a coordination game. There are two possible Nash Equilibria with pure strategies $(C, D)$ and $(D, C)$.
    \end{enumerate}
    
    Moving backwards, consider the regulator's decision. Recall that
    R maximizes social utility. Again, three scenarios emerge:
   \begin{enumerate}
        \item $x < 2 \left[ \frac{P-S}{T}\right]$. The social utility is $P$.
        
        \item $x > 2 \left[ 1- \frac{R}{T}\right]$. The social utility is $R$.
        
        \item $x \in \left( 2 \left[ \frac{P-S}{T}\right], 2 \left[ 1- \frac{R}{T}\right] \right) $. The social utility is $\frac{S+T}{2}$.
    \end{enumerate}
    
    As $R > P$ and $R > \frac{T+S}{2}$ (as requested in the IPD),
    the regulator maximizes his payoff choosing $x > 2 \left[ 1- \frac{R}{T}\right]$. Therefore, $(x, C, C)$, with $x > 2 \left[ 1- \frac{R}{T}\right]$ is a subgame perfect equilibrium, and 
    we can foster cooperation in the one-shot version of the game.
    
\section{One-shot game for the centralized case plus incentives}\label{sec:oneshot_inc}

Under this scenario, we consider the reward bimatrix in Table \ref{tab:payoffIPD_inc}, 
where $I$ is the incentive introduced by the Regulator. 

 \begin{table}[h!]
    \begin{center}
    \begin{tabular}{cl|lll}
    \multicolumn{1}{l}{}                                   &     & \multicolumn{3}{l}{\textbf{DDO}} \\ \cline{3-5} 
    \multicolumn{1}{l}{}                                   &     & $C$         &       & $D$        \\ \hline
    \multicolumn{1}{c|}{\textbf{Citizen}} & $C$ & $R+I,R+I$       &       & $S,T$      \\
    \multicolumn{1}{c|}{}                                  &     &             &       &            \\
    \multicolumn{1}{c|}{}                                  & $D$ & $T,S$       &       & $P,P$     
    \end{tabular} 
    \end{center}
    \caption{Utilities for the data sharing game with incentives}
    \label{tab:payoffIPD_inc}
    \vspace{-2ex}
    \end{table}

Consider the case in which the agents take 
the $(C, D)$ pair of actions. In this case, they perceive rewards $(S, T)$.
After tax collection and distribution, it leads to
$(S - \frac{Sx}{2} +  \frac{Tx}{2}, T -  \frac{Tx}{2} +  \frac{Sx}{2})$, with $x$ being the tax rate collected by the Regulator. In order to ensure 
that $(C,C)$ is a Nash equilibrium, two conditions must hold:
\begin{itemize}
    \item $S - \frac{Sx}{2} +  \frac{Tx}{2} > P$, so that agents do not 
    switch from $(C,D)$ to $(D,D)$. This simplifies to $x > 2\frac{(P-S)}{T-S}$.
    \item $R + I > T -  \frac{Tx}{2} +  \frac{Sx}{2}$, so that 
    the agents do not switch from $(C, C)$ to $(C,D)$. This simplifies to $x > 2\frac{T - (R+I)}{T-S}$.
\end{itemize}

This shows that even if the gap between $T$ and $R$ is large, with the
aid of incentives both agents could reach mutual cooperation, also under a tax framework, since $I$ can grow arbitrarily to ignore the second restriction.
\pagebreak 

\end{document}